\documentclass[pre,twocolumn,superscriptaddress,showpacs]{revtex4}

\usepackage{amsmath}
\usepackage{graphicx}

\begin{document}

\title{Nonperiodic echoes from quantum mushroom billiard hats}

\author{B.~Dietz}
\affiliation{Institut f{\"u}r Kernphysik, Technische Universit{\"a}t Darmstadt, D-64289 Darmstadt, Germany}

\author{T.~Friedrich}
\affiliation{Institut f{\"u}r Kernphysik, Technische Universit{\"a}t Darmstadt, D-64289 Darmstadt, Germany}
\affiliation{GSI Helmholtzzentrum f{\"u}r Schwerionenforschung GmbH, D-64291 Darmstadt, Germany}

\author{M.~Miski-Oglu}
\affiliation{Institut f{\"u}r Kernphysik, Technische Universit{\"a}t Darmstadt, D-64289 Darmstadt, Germany}

\author{A.~Richter}
\email{richter@ikp.tu-darmstadt.de}
\affiliation{Institut f{\"u}r Kernphysik, Technische Universit{\"a}t Darmstadt, D-64289 Darmstadt, Germany}
\affiliation{$\rm ECT^*$, Villa Tambosi, I-38100 Villazzano (Trento), Italy}

\author{F.~Sch{\"a}fer}
\affiliation{Institut f{\"u}r Kernphysik, Technische Universit{\"a}t Darmstadt, D-64289 Darmstadt, Germany}

\author{T.~H.~Seligmann}
\affiliation{Instituto de Ciencias F\'isicas, Universidad Nacional Aut\'onoma de M\'exico, 
Chamilpa, Cuernavaca Morelos, M\'exico}
\affiliation{Instituto de Ciencias F\'isicas, 
Centro Internacional de Ciencias A.C. C.U., Chamilpa, Cuernavaca Morelos, M\'exico}

\date{\today}

\begin{abstract}
Nonperiodic tunable quantum echoes have been observed in experiments with an open
microwave billiard whose geometry under certain conditions provides Fibonacci like sequences of classical delay times. 
These sequences combined with the reflection at the opening induced by the wave character 
of the experiment and the size of the opening 
allow to shape quantum pulses. The pulses are obtained by response of an integrable scattering 
system.

\end{abstract}


\maketitle Wave mechanics in billiards can be implemented by flat microwave
cavities~\cite{1,2,3,4}. They model single particle quantum aspects of
mesoscopic structures, e.g.\ two dimensional electron gases in quantum dots or
more complicated systems~\cite{5,6,7,8}. Recently billiard systems with large
openings have attracted attention~\cite{9,10,11,12,13,14}. In \cite{18} properties 
of the classical dynamics of open mushroom billiards were investigated. 
Mushroom billiards, which were proposed by Bunimovich \cite{18a} recently, 
consist of one or more circular or elliptic hats which are attached to stems 
composed of straight walls. Their phase space has the particular property that 
the chaotic and the regular areas are sharply separated with no fractal 
structure in the border region. The mushroom billiards considered in \cite{18} 
consist of a circular hat and a triangular stem. Trajectories of particles 
passing the stem belong to the chaotic part of phase space, whereas the hat 
comprises chaotic and regular trajectories \cite{18a,18b,18c}. In \cite{18} we were interested in
the number of bounces a particle entering the hat of the mushroom experiences before 
exiting back into the stem. Note that the starting conditions for these 
trajectories all belong to the chaotic part of the phase space. The result of 
these studies of the classical dynamics was that for a fixed angular momentum 
a selective number of bounces, in fact a total of three, is possible. Only trajectories
of particles, which are scattered from inside the stem immediately into the hat are considered and 
followed only until they 
reenter the stem. Thus, as these particles never touch the boundary of the stem the results presented
in \cite{18} are independent of its geometry. Indeed it was shown there, that the selectivity persists
when considering the hat as an open scattering system with the opening obtained by 
removing the stem. Then one observes for a fixed angular momentum only three different 
delay times, i.e. times a particle sent into the hat spends there before exiting it.  These
observations lead to the question how far this selectivity influences the pulse structure of 
the corresponding 
open quantum billiard. Here we report on
measurements of the time resolved response of such a scattering system to
incident waves, and the detection of aperiodic and selective pulse
sequences. In the experiments a pulse is sent into an open mushroom billiard hat from the outside and the pulses sent back to the exterior are recorded. The
sensitive dependence of the corresponding classical response on the size of
the opening can be used as a guideline to design desirable pulse sequences in
the wave domain. This is in stark contrast to systems whose classical or ray
dynamics is mixed with no sharply divided phase space or chaotic. In the former case 
periodic echo signals were
seen~\cite{15} and theoretically understood~\cite{16,17} for the short time
behavior, in the latter a noisy response is expected. The experiment described here was
performed for a quantum billiard with the shape of a quarter circle, but the
flexibility of design can be enhanced e.g.\ by deforming the circle to an
ellipse~\cite{18}. Keeping in mind the analogy to open nano structures e.g.\
for two dimensional electron gases~\cite{19,20,21} this paves the way to
convert a simple quantum pulse into a complicated non-periodic pulse sequence
and thus to obtain a tunable quantum pulse generator by a simple scattering
mechanism.

\begin{figure}[ht]
	\centering
	\includegraphics[width=8cm]{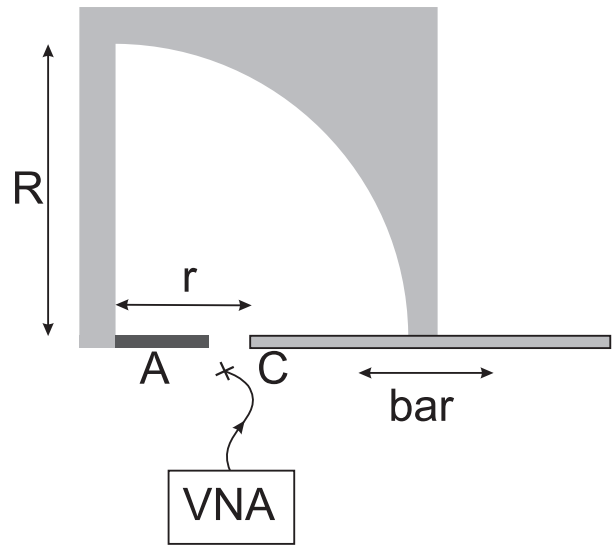}
	\caption{Sketch of the experimental setup (top view). The inner white part
	of the light gray area indicates the quarter circle shape of the flat
	microwave resonator. The radius is $R$ and the size of the adjustable
	opening is $r$. An antenna ($\times$) in front of the opening near point C
	(edge of the bar) couples the microwave signal in and out and is attached to
	a vector network analyzer (VNA). The dark gray bar A indicates microwave
	absorber material inserted into a part of the opening of the cavity.}
	\label{fig:img1}
\end{figure}

We recall briefly that billiards, which are paradigmatic dynamical systems~\cite{22,23}, are
two-dimensional domains with free motion except for specular reflections at
the walls. The corresponding quantum systems are determined by the time
independent Schr\"odinger equation with Dirichlet boundary conditions at the walls. For the experimental 
investigation of such quantum billiards we exploit the equivalence of the related Schr{\"o}dinger 
equation and of the Helmholtz equation for the electric field strength in a flat, cylindrical microwave 
resonator of corresponding shape below that frequency, where the first transversal electric mode is
excited \cite{23a,23b}. Up to this frequency the electric field strength is perpendicular to the top 
and bottom plate of the resonator and 
the Helmholtz equation is scalar. Such a microwave resonator consists of
two parallel plates and a third plate with a hole of the shape of the billiard squeezed in between.
Specifically, the hole has the form of a quarter circular boundary with radius $R
= 240~{\rm mm}$ such that the resonator is open along one straight line of the
quarter circle. Note that this geometry is a realization of a desymmetrized open mushroom
billiard with circular hat~\cite{24,25}. The
size $r$ of the hole can be adjusted by a bar to yield the shape indicated in
Fig.~\ref{fig:img1}. The separation of the parallel plates is 5~mm. Microwave power was emitted into 
the resonator and received by the same dipole antenna. A Vector Network Analyzer (VNA) provided the
rf signal for frequencies between 1--17~GHz well within the limit for the validity of the scalar 
Helmholtz equation, the excitation frequency being increased with a step size of 50~kHz. The VNA measured
the ratio of the received to the emitted microwave power and the relative phase thus yielding the complex 
scattering matrix elements for the scattering of electromagnetic waves from the antenna into the resonator 
and back to it. The antenna was placed perpendicular to the billiard plates in 7~mm distance from 
the opening outside the cavity and 15~mm from corner C of the bar, cf.\ Fig.~\ref{fig:img1}.
We introduced microwave absorbing material A along the opening of the
billiard up to a distance of 15~mm to the antenna thereby damping multiple
wave reflections at the opening. The measurements were performed 
for different opening parameters $r/R$ of the billiard.

\begin{figure}[ht]
	\centering
	\includegraphics[width=8cm]{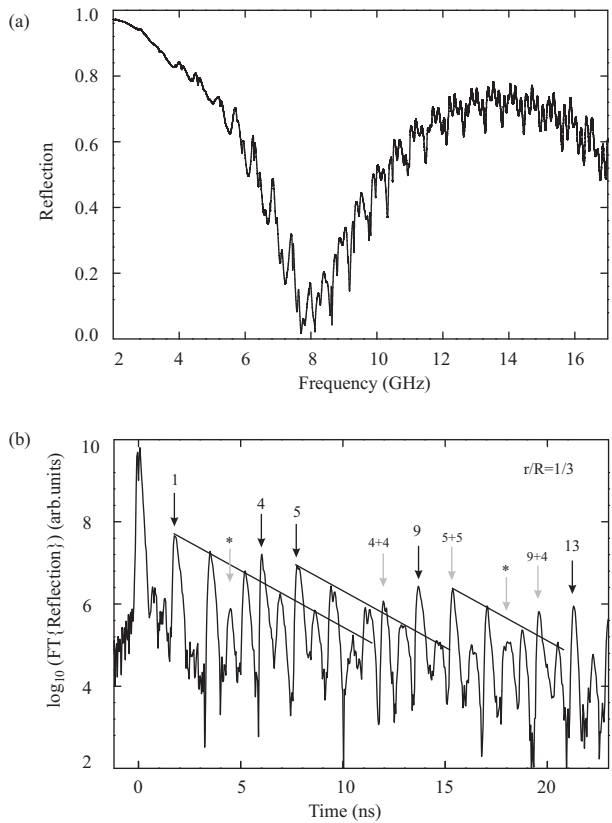}
	\caption{Upper panel: Reflection spectrum at the antenna shown in
	Fig.~\ref{fig:img1} (linear scale). Lower Panel: Fourier transform of the
	spectrum in semilog-scale. The black arrows mark times at which classical
	echoes occur, whereas gray arrows mark times of trajectories facilitated by
	quantum effects. Bounce numbers corresponding to these times are also shown.
	Exponentially decaying peak sequences of
	equal spacings are connected by straight lines. Some peaks correspond to a
	combination of bounce numbers ($4+4$, e.g.) or more complex systematics
	($\ast$).}
	\label{fig:img2}
\end{figure}

The overall shape of the reflection spectra shows a minimum close to 8 GHz due
to the emission characteristics of the antenna, c.f.\ upper panel of
Fig.~\ref{fig:img2}. The scattering information we are interested in is
contained in the fine structure imprinted on this overall shape. A Fourier
transform of the entire spectrum yields the response to a short pulse in the
time domain. The modulus square of the signal obtained in this way decays 
as a power law with decay
exponent $\gamma\simeq 1.95$. This value is very close to the predicted   
one~\cite{26} of 2 for classical particles which escape from a billiard with
integrable dynamics. However, here we are
interested in the short time characteristics of the system, which deviates
from this behavior. The lower panel of Fig.~\ref{fig:img2} shows the time
response of the open quarter circle with $r/R = 1/3$ for short times. An
aperiodic sequence of peaks is clearly visible, and their strengths decay on
average with increasing time. The peak seen near time 0 is related to the
smooth frequency dependence of the emission characteristics of the antenna 
observed as a broad dip in Fig.~\ref{fig:img2}.

As mentioned above it was shown in~\cite{18} that in the corresponding classical 
scattering system a particle 
injected from the outside into the quarter circle billiard encounters only a certain 
number of reflections on the circular boundary (shortly called bounces) from a scarce set of
possible numbers before it leaves it. For the opening ratio $r/R = 1/3$ the sequence of
allowed bounce numbers $n$ is $1, 4, 5, 9, 14, 23, 37, 51, \ldots$. It
proceeds up to the number 37 like a generalized Fibonacci series, i.e.\ each
number is given by the sum of the two previous ones. For larger $n$ the sequence is no longer
Fibonacci like but still each occurring number is the sum of two smaller ones. When analyzed in detail~\cite{18}
one finds that this sum rule is strictly obeyed within finite intervals of
angular momentum values, i.e.\ of the impact parameter with
respect to the center of the circle. Each interval is bordered by two singularities
of diverging bounce numbers resulting from the existence of parabolic
manifolds~\cite{27}. This behavior can be explained in terms of
number theoretical properties of the circle map~\cite{28}.

For an interpretation of the peaks in the Fourier transformed experimental
spectra in terms of the bounce numbers $n$ of the classical scattering dynamics these need to
be converted into time delays. Using conservation of the angular momentum, respectively
the impact parameter $b$, the time $T$ a particle needs for $n$ reflections
at the circular boundary is given as
\begin{equation}
	T = 2\, R\, n\, \sqrt{1 - (b/R)^2}/v\, .
	\label{eqn:1}
\end{equation}
Here $v$ denotes the velocity of the particle. Note that the possible values
for $b$ are restricted by the size of the opening, and typically $b \approx 0$
for $n = 1$ and $b \approx r$ for large bounce numbers. Applying this formula
(with $v = c$ the speed of light and of the microwave propagation) to the
sequence of bounce numbers given above we obtain classical predictions for the possible delay times, i.e. the positions of the echoes for the opening ratio $r/R=1/3$. We mark these
times with black arrows identified by the corresponding bounce numbers $n$ in
the lower panel of Fig.~\ref{fig:img2}, and indeed find that they coincide
with the dominant peaks of the measured response, i.e.\ in a microwave
experiment mimicking an open quantum billiard of corresponding shape we detected 
scattering echoes at the predicted times.

However, many additional peaks appear although the predicted peaks protrude
above the average decay of the time response. The smaller peaks marked by stars
are related to
multiple reflections caused by diffraction at the edges of the opening. The peaks
corresponding to 1 and 5 bounces at the circular boundary as well as the one
denoted by $5+5$ are followed by an exponentially decaying sequence of peaks
with constant spacing, equal in all three cases. To guide the eye, the
exponential decay is indicated by the straight lines in the lower panel of
Fig.~\ref{fig:img2}. We see that also the decay rates coincide. The sequences
are caused by waves that hit the corners of the opening and there get
partially scattered into the 1-bounce orbit. Both escape and reflection of the
1-bounce orbit at the opening may happen repeatedly, leading to the
exponentially decaying sequence of peaks. The purpose of the absorbing materials covering part of the
opening is to suppress 1-bounce reflections. Experimental setups without the
absorber material and with the antenna nearer to the center showed dominance
of trivial 1-bounce peaks and many secondary peaks for large openings. Other peaks can be attributed to a wave impinging on the opening and scattered
into classical orbits which bounce more than once at the circular boundary
before they hit the opening again. These peaks are marked by gray arrows and
are labeled by $n+m$ for the scattering of an $n$ bounce orbit into an $m$ bounce
orbit. Effects caused by
the coupling to regular modes of waves reentering the cavity due to diffraction at
the corners of the opening were investigated in detail in \cite{28a}, of refracted fields
reentering a dielectric cavity with the shape of a mushroom in \cite{28b}.

With the argumentation given above we understand the most prominent peaks of the response
function, but we may ask why we do not see the exponential decay after the 4
and the 9-bounce peaks. The reason is that the sequence of peaks following the
4 (9) bounce peak coincides with that of the 5 ($5+5$) bounce peaks within the
peak width determined by the spectral range and are thus superimposed and then
decay, as mentioned above, exponentially in a sequence of repeated single
bounces. A sharp eye might detect the double peaks in these sequences, but
this effect is at the limit of our resolution. Other smaller peaks, e.g.\
those labeled by the stars, admit more complicated assignments but are still
understood in terms of diffractive orbits. Summarizing the analysis of
Fig.~\ref{fig:img2} we conclude that the pulse structure is essentially
determined by the possible classical escape times (depending on the hole
size), by diffractions at the opening (depending on the usage of microwave
absorbing material) and by the coupling strength of the waves to the internal
states corresponding to classical structures (depending on the antenna
position). Modifying the experimental setup and thereby changing any or all of
the three determining factors we can drastically change the output pulse
sequence.

\begin{figure}[ht]
	\centering
	\includegraphics[width=8cm]{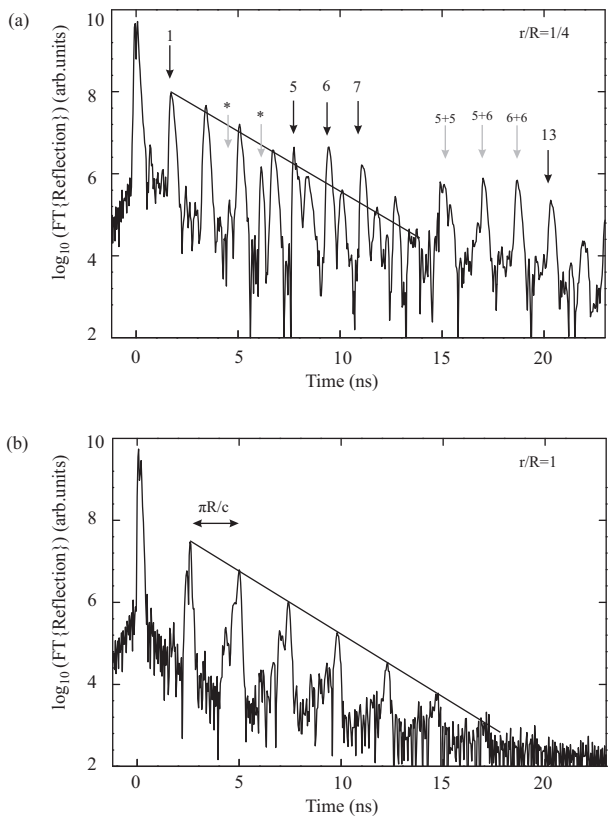}
	\caption{Same as the lower panel of Fig.~\ref{fig:img2} for opening sizes
	$r/R = 1/4$ (upper panel) and $r/R = 1$ (lower panel), where we find
	whispering gallery dynamics. The time between two peaks is $\pi\,R/c$.}
	\label{fig:img3}
\end{figure}

To support the validity of those conclusions we performed measurements for
different opening ratios $r/R$, that is allowed classical escape times. In all
cases the interpretation as given above explains the detected echoes or pulse
sequences. In the upper panel of Fig.~\ref{fig:img3} we show the time response
for $r/R = 1/4$. Note that the Fibonacci like behavior is not as prominent as
in the former case, as already for $n=7$ the number of bounces does not equal the sum of the previous two possible bounce numbers, but that of the first and the third one. However, the sequence starting with $n = 1$ is
characterized by a fixed period, and the superposition of several of these
sequences again leads to a highly aperiodic pulse signal. Finally we see in
the lower panel of Fig.~\ref{fig:img3} that in the geometry of the quarter
circle billiard periodic echoes can also be realized, though they might be of
less interest. This is feasible for $r/R = 1$, i.e.\ for the fully opened
billiard. In this case the wave travels along high order polygonal orbits
which follow the circular boundary closely, known as whispering gallery
orbits~\cite{29}. The time between two successive echoes is the time
needed to travel forth and back along the circular boundary at the speed of
light.

Previous work~\cite{15} has shown, that the time structure of
quantum signals will reflect certain classical properties even if the
experiment is carried out far from the classical limit. Yet known examples
lead to periodic or near periodic response. Our experiment provides an example
for a pulse response in terms of aperiodic echoes by wave scattering off an
integrable system. Deviations from the time structure of the classical problem
are mainly due to diffraction at the opening. Moreover, in our microwave experiment
the quasi two dimensional interior of the
billiard is coupled to the three dimensional free space. Such effects are well understood
and can be controlled. Though microwave billiards as model systems neglect the many body
character of a quantum dot as well as charges and spins of the electrons, the
detected aperiodicity is expected to be visible also in ballistic scattering
experiments on the nano scale. As the time resolved treatment of transport
through quantum dots was extremely successful in the last decade both
theoretically~\cite{21,30,31} and experimentally~\cite{32,33,34,35}, the
future development of quantum pulse generators, providing a large diversity of
pulse responses seems feasible.

\begin{acknowledgments}
This work was supported by Deutsche Forschungsgemeinschaft within SFB
634 and by PAPIT (UNAM) projects No. IN-111607 (DGAPA) and 79988 (CONACyT). 
F.S.\ got a grant from Deutsche Telekom Foundation.
\end{acknowledgments}

\end{document}